\documentclass[pre,aps,twocolumn,superscriptaddress,amssymb,amsmath,floatfix]{revtex4-2}

\usepackage{graphicx}
\usepackage[hidelinks]{hyperref}
\usepackage{xcolor}
\usepackage{booktabs}
\usepackage{tikz}
\usetikzlibrary{arrows.meta,angles,quotes,calc,positioning}
\usepackage{bm}

\newcommand{\e}{\mathrm{e}}

\newcommand{\av}[1]{\left\langle#1\right\rangle}
\newcommand{\Tr}{\operatorname{Tr}}

\definecolor{agentblue}{RGB}{45,100,180}
\definecolor{agentorange}{RGB}{225,125,35}
\definecolor{friendgreen}{RGB}{35,145,90}
\definecolor{foered}{RGB}{200,65,65}
\definecolor{nematicpurple}{RGB}{125,80,170}
\definecolor{neutralgray}{RGB}{120,120,120}

\begin{document}

\title{When Does Triadic Balance Reduce to Pairwise Order?}

\author{Sepideh Arab Mohammadi}
\affiliation{Department of Physics, Shahid Beheshti University, Evin, Tehran 1983969411, Iran}

\author{Amir Kargaran}
\affiliation{Department of Physics, Shahid Beheshti University, Evin, Tehran 1983969411, Iran}

\author{Ali Hosseiny}
\email{alihd22@gmail.com}
\affiliation{Department of Physics, Shahid Beheshti University, Evin, Tehran 1983969411, Iran}

\author{Reza Jafari}
\affiliation{Department of Physics, Shahid Beheshti University, Evin, Tehran 1983969411, Iran}

\author{Abbas Ali Saberi}
\email{asaberi@constructor.university}
\affiliation{School of Science, Constructor University, Campus Ring 1, 28759 Bremen, Germany}

\date{November 1, 2026}

\begin{abstract}
Structural-balance interactions are naturally defined on triads, but when pair affinities are themselves generated by continuous node attributes, it is not clear whether the resulting interaction remains genuinely higher order. Here we address this question for agents represented by unit vectors in an attribute space, with pair affinity given by their scalar product and a Heider product assigned to each triangle. For planar attributes, we show that the triadic interaction reduces exactly, at finite system size, to a pairwise Curie--Weiss XY model of the doubled angles. The ordered phase is therefore nematic and undergoes a continuous mean-field $O(2)$ transition at $k_{\mathrm B}T_c=J/8$. The same reduction reveals a local $\mathbb Z_2$ degeneracy: each nematically ordered state corresponds to $2^N$ polar assignments, whose binomial multiplicities make two nearly equal opposite groups typical at low temperature. Monte Carlo simulations support the exact critical temperature, the mean-field Binder-cumulant value, and the predicted complete-graph finite-size scaling. Finally, a tensor representation of the interaction shows that the planar reduction is dimension specific: in two dimensions the cubic invariant is constrained by lower-order invariants, whereas for $d\geq3$ it is generically independent. These results identify when attribute-induced triadic balance is effectively pairwise and when genuinely higher-order structure can remain.
\end{abstract}

\maketitle

\section{Introduction}

Heider's balance theory organizes signed relations through a constraint on triangles: a triad is balanced when the product of its three link signs is positive \cite{heider1,heider2}. On a complete graph this condition leads to either an all-positive ``paradise'' state or a partition into two internally friendly and mutually antagonistic factions \cite{cartwright,antal1,antal2}. Large signed-network data sets have provided empirical tests of balance and related signed-triad patterns \cite{szell,facchetti,leskovec,allahyari}. Statistical-physics formulations promote imbalance to an energy and use stochastic or thermal dynamics to study ordering, metastability, and phase transitions. Thermal Heider models with link degrees of freedom can exhibit discontinuous balance--disorder transitions, whereas coupling node and link variables can instead produce a continuous transition \cite{kargaran2021coevolutionary}. Recent extensions to multiplex, Erd\H{o}s--R\'enyi, and planar networks further show how balance transitions depend on interlayer coupling, connectivity, and local triangle topology \cite{malarz2022meanfield,mohandas2024multiplex,mohandas2025paradise,malarz2025archimedean}. A recent complementary result shows that, within the structurally balanced sector, binary Heider links can be represented by products of Ising spins on complete and arbitrary graphs \cite{burda2026ising}. That correspondence concerns binary link variables subject to structural balance, whereas the reduction derived below is a configuration-by-configuration identity for continuous affinities generated by node attributes.

A distinct but complementary question concerns the origin of the links themselves. Continuous-link formulations allow the strength of friendship or hostility to vary \cite{kulakowski,marvel2}, while homophily-based models generate relations from similarities between multidimensional attributes or opinions. Attribute-based balance models explicitly couple triadic balance to agent features \cite{gorski2020homophily,pham2021balance}; multidimensional attributes have also been used to expose signatures of Heider dynamics in empirical and model networks \cite{linczuk2023multidimensional,oloomi2021,masoumi2021}; dyadic homophily alone can reproduce characteristic triad statistics \cite{pham2022triad}; and multidimensional spin descriptions have been used to study homophilic group formation \cite{korbel2023homophily}. {
Agbanusi and Bronski studied an energy-based model with continuous node positions and edge opinions, obtaining consensus and balanced two-group states \cite{agbanusi2018balance}.
} These results motivate a sharper structural question: \emph{when affinities are induced by continuous node attributes but the Hamiltonian is imposed on triads, does the apparent three-body interaction contain an irreducible higher-order degree of freedom?}

{
Here $d=2$ means that each agent carries a two-component attribute vector; it does not specify the geometry of the underlying network. We take the affinity to be the scalar product of two attribute vectors, a definition that is valid in any attribute dimension. For $d=2$, the triangle energy can be rewritten exactly as a constant plus interactions between pairs of doubled angles. On a complete graph these pair interactions are uniform and define the Curie--Weiss XY model, the fully connected model of planar spins \cite{stanley1968}. Ordering of the doubled angles corresponds to a nematic state: an attribute vector and its opposite contribute identically to the order parameter.

The exact reduction applies to any specified collection of triangles, although the ensuing critical behavior depends on the network. We use the complete graph for an exact thermodynamic solution and then examine the dimensional dependence of the interaction separately.
}

\section{Model}

We consider $N$ agents on a fully connected graph. {
The network specifies which agents form triangles, while the attribute-space dimension $d$ counts the components of an agent's vector; these are independent notions.
} The internal state of agent $i$ is a unit vector in a two-dimensional interest space,
\begin{equation}
    \mathbf S_i=(\cos\theta_i,\sin\theta_i),
\end{equation}
where $\theta_i\in[0,2\pi)$ specifies the orientation of the agent in the attribute space. The two axes may represent any pair of normalized interests, opinions, or attributes; only relative orientations enter the model.

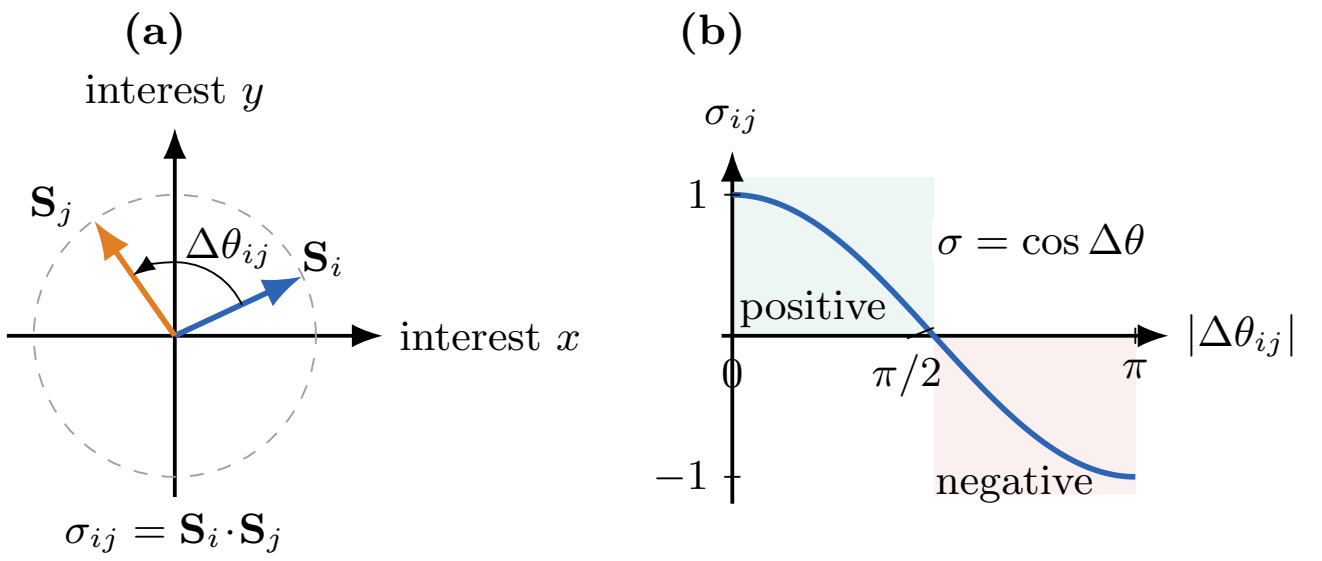
\begin{figure}[tbp]
\centering
\resizebox{\columnwidth}{!}{%
\begin{tikzpicture}[font=\small,>=Latex]
    \begin{scope}[shift={(0,0)}]
        \node[text=black,font=\bfseries] at (-0.15,2.25) {(a)};
        \draw[->,black,thick] (-1.25,0) -- (1.55,0) node[right,text=black]{interest $x$};
        \draw[->,black,thick] (0,-1.20) -- (0,1.55) node[above,text=black]{interest $y$};
        \draw[neutralgray!70,dashed] (0,0) circle (1.05);
        \draw[->,agentblue,very thick] (0,0) -- ({1.05*cos(25)},{1.05*sin(25)});
        \draw[->,agentorange,very thick] (0,0) -- ({1.05*cos(125)},{1.05*sin(125)});
        \node[text=black,anchor=west] at (0.83,0.57) {$\mathbf S_i$};
        \node[text=black,anchor=east] at (-0.62,0.96) {$\mathbf S_j$};
        \draw[->,black] (0.50,0.23) arc[start angle=25,end angle=125,radius=0.55];
        \node[text=black] at (0.4,0.65) {$\Delta\theta_{ij}$};
        \node[text=black,align=center] at (0,-1.47) {$\sigma_{ij}=\mathbf S_i\!\cdot\!\mathbf S_j$};
    \end{scope}

    \begin{scope}[shift={(4.15,0)}]
        \node[text=black,font=\bfseries] at (-0.15,2.25) {(b)};
        \fill[friendgreen!8] (0,0) rectangle (1.50,1.18);
        \fill[foered!8] (1.50,-1.18) rectangle (3.00,0);
        \draw[->,black,thick] (0,-1.25) -- (0,1.38) node[above,text=black] {$\sigma_{ij}$};
        \draw[->,black,thick] (-0.08,0) -- (3.25,0) node[right,text=black] {$|\Delta\theta_{ij}|$};
        \draw[agentblue,very thick,domain=0:180,samples=100]
            plot ({\x/60},{1.05*cos(\x)});
        \draw[black] (0,0.06)--(0,-0.06) node[below,text=black] {$0$};
        \draw[black] (1.5,0.06)--(1.3,-0.02) node[below,text=black] {$\pi/2$};
        \draw[black] (3.0,0.06)--(3.0,-0.06) node[below,text=black] {$\pi$};
        \draw[black] (0.06,1.05)--(-0.06,1.05) node[left,text=black] {$1$};
        \draw[black] (0.06,-1.05)--(-0.06,-1.05) node[left,text=black] {$-1$};
        \node[text=black] at (0.6,0.2) {positive};
        \node[text=black] at (2.1,-1.1) {negative};
        \node[text=black,fill=white,inner sep=1pt] at (2.30,0.72) {$\sigma=\cos\Delta\theta$};
    \end{scope}
\end{tikzpicture}%
}

\caption{Continuous affinity generated by planar attributes. (a) Each agent is represented by a unit vector, and only the angular separation $\Delta\theta_{ij}=\theta_i-\theta_j$ matters. (b) The scalar-product affinity varies continuously from $+1$ to $-1$ as the separation increases, rather than being assigned as an independent binary link variable.}
\label{fig:schematic}
\end{figure}

The affinity between agents $i$ and $j$ is defined by
\begin{equation}
    \sigma_{ij}=\mathbf S_i\cdot\mathbf S_j
    =\cos(\theta_i-\theta_j),
    \label{eq:affinity}
\end{equation}
so that $\sigma_{ij}\in[-1,1]$. Parallel vectors have maximal positive affinity, orthogonal vectors are neutral, and antiparallel vectors have maximal negative affinity. Figure \ref{fig:schematic} shows both the geometry and the continuous dependence of affinity on angular separation.

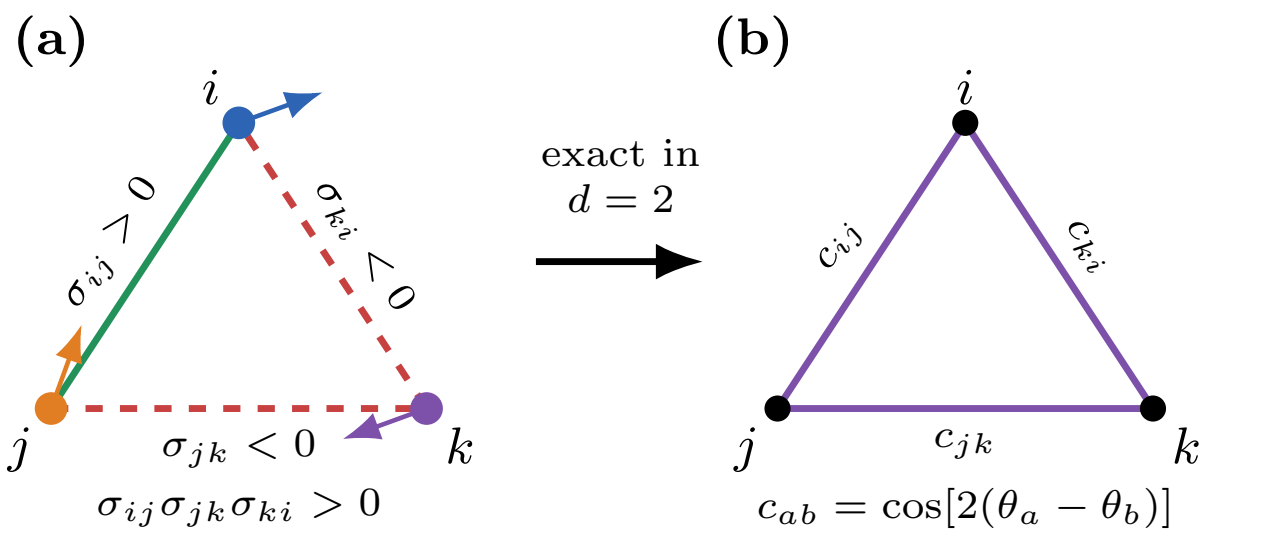
\begin{figure}[tbp]
\centering
\resizebox{0.94\columnwidth}{!}{%
\begin{tikzpicture}[font=\small,>=Latex]
    \node[text=black,font=\bfseries] at (-2.55,1.60) {(a)};
    \coordinate (I) at (-1.40,1.10);
    \coordinate (J) at (-2.55,-0.65);
    \coordinate (K) at (-0.25,-0.65);
    \draw[friendgreen,very thick] (I)--(J)
      node[midway,sloped,above,text=black,font=\scriptsize] {$\sigma_{ij}>0$};
    \draw[foered,very thick,dashed] (J)--(K)
      node[midway,below,text=black,font=\scriptsize] {$\sigma_{jk}<0$};
    \draw[foered,very thick,dashed] (K)--(I)
      node[midway,sloped,above,text=black,font=\scriptsize] {$\sigma_{ki}<0$};
    \fill[agentblue] (I) circle (0.10); \node[text=black,above left] at (I) {$i$};
    \fill[agentorange] (J) circle (0.10); \node[text=black,below left] at (J) {$j$};
    \fill[nematicpurple] (K) circle (0.10); \node[text=black,below right] at (K) {$k$};
    \draw[->,agentblue,thick] (I) -- ++(20:0.55);
    \draw[->,agentorange,thick] (J) -- ++(70:0.55);
    \draw[->,nematicpurple,thick] (K) -- ++(200:0.55);
    \node[text=black,font=\scriptsize] at (-1.40,-1.26)
       {$\sigma_{ij}\sigma_{jk}\sigma_{ki}>0$};

    \draw[->,black,very thick] (0.42,0.25) -- (1.45,0.25);
    \node[text=black,align=center,font=\scriptsize] at (0.94,0.78)
       {exact in\\$d=2$};

    \node[text=black,font=\bfseries] at (1.75,1.60) {(b)};
    \coordinate (Ip) at (3.05,1.10);
    \coordinate (Jp) at (1.90,-0.65);
    \coordinate (Kp) at (4.20,-0.65);
    \draw[nematicpurple,very thick] (Ip)--(Jp)
      node[midway,sloped,above,text=black,font=\scriptsize] {$c_{ij}$};
    \draw[nematicpurple,very thick] (Jp)--(Kp)
      node[midway,below,text=black,font=\scriptsize] {$c_{jk}$};
    \draw[nematicpurple,very thick] (Kp)--(Ip)
      node[midway,sloped,above,text=black,font=\scriptsize] {$c_{ki}$};
    \fill[black] (Ip) circle (0.08); \node[text=black,above] at (Ip) {$i$};
    \fill[black] (Jp) circle (0.08); \node[text=black,below left] at (Jp) {$j$};
    \fill[black] (Kp) circle (0.08); \node[text=black,below right] at (Kp) {$k$};
    \node[text=black,font=\scriptsize] at (3.05,-1.26)
       {$c_{ab}=\cos[2(\theta_a-\theta_b)]$};
\end{tikzpicture}%
}

\caption{Local structure behind the exact reduction. (a) A balanced Heider triad generated by continuous affinities, illustrated by one positive and two negative pair relations. The arrows at the nodes indicate the underlying attribute directions. (b) For planar attributes, the same triad contribution obeys $\tau_{ijk}=\tfrac14(1+c_{ij}+c_{jk}+c_{ki})$, with $c_{ab}=\cos[2(\theta_a-\theta_b)]$. Thus the triadic product becomes pairwise in the doubled angles before any thermodynamic or mean-field approximation is made.}
\label{fig:triangle}
\end{figure}

Following structural-balance theory, the energy of a triangle $(i,j,k)$ is
{
\begin{equation}
\tau_{ijk}\equiv\sigma_{ij}\sigma_{jk}\sigma_{ki},\qquad
E_{ijk}=-J\tau_{ijk},\quad J>0.
\end{equation}
For signed links, $\tau_{ijk}>0$ corresponds to either three friendly relations or one friendly and two antagonistic relations; $\tau_{ijk}<0$ corresponds to an imbalanced triad. For continuous affinities the same product gives a continuous, gauge-invariant energy preference rather than a binary balance label.
} The full Hamiltonian is
\begin{equation}
    H_N=-\frac{J}{N^2}\sum_{i<j<k}
    \sigma_{ij}\sigma_{jk}\sigma_{ki},
    \label{eq:Htriad}
\end{equation}
where the $N^{-2}$ Kac scaling makes the energy extensive on the complete graph. Figure \ref{fig:triangle} illustrates a balanced triangle and previews the exact planar reduction derived below.

The Hamiltonian has two exact symmetries. First, a global rotation
\begin{equation}
    \theta_i\rightarrow\theta_i+\alpha \qquad \forall i
\end{equation}
leaves every relative angle unchanged and gives a global $U(1)$ symmetry. Second, it is invariant under an independent $\pi$ shift of any one node,
\begin{equation}
    \theta_i\rightarrow\theta_i+\pi.
    \label{eq:localZ2}
\end{equation}
Under Eq. \eqref{eq:localZ2}, every affinity incident on $i$ changes sign. Each triangle containing $i$ contains exactly two such affinities, so the signs cancel in their product. The resulting local $\mathbb Z_2$ symmetry makes the polar magnetization {
$m_{\rm p}=N^{-1}\left|\sum_i\e^{\mathrm{i}\theta_i}\right|$, which changes under independent node flips,
} inappropriate as an order parameter and instead selects the nematic variable
\begin{equation}
    q=\frac{1}{N}\sum_{i=1}^N \e^{2\mathrm{i}\theta_i},
    \qquad n=|q|.
    \label{eq:nematic}
\end{equation}

\section{Exact reduction and mean-field solution}
\label{Analytical}

\subsection{Exact reduction of the triadic interaction}

For one triangle, define $A=\theta_i-\theta_j$ and $B=\theta_j-\theta_k$. Since $\theta_k-\theta_i=-(A+B)$,
\begin{align}
\cos A\cos B\cos(A+B)
=\frac14\big[1+\cos2A+\cos2B \nonumber\\
\hspace{3.0cm}+\cos2(A+B)\big].
\end{align}
This gives
\begin{align}
\tau_{ijk}=\frac14\big[1
&+\cos2(\theta_i-\theta_j)
+\cos2(\theta_j-\theta_k) \nonumber\\
&+\cos2(\theta_k-\theta_i)\big].
\label{eq:triadidentity}
\end{align}
Because Eq.~\eqref{eq:triadidentity} is a local identity, the pairwise reduction itself does not require a complete graph. For an arbitrary collection $\mathcal T$ of triangles with couplings $J_{ijk}$,
\begin{align}
-\sum_{(ijk)\in\mathcal T}J_{ijk}\tau_{ijk}
={}&-\frac14\sum_{(ijk)\in\mathcal T}J_{ijk} \nonumber\\
&-\frac14\sum_{i<j}J^{\rm eff}_{ij}
\cos\left[2(\theta_i-\theta_j)\right],
\label{eq:generalgraph}
\end{align}
where $J^{\rm eff}_{ij}=\sum_{k:(ijk)\in\mathcal T}J_{ijk}$. Thus, for planar attributes, any interaction built as a sum of these triangle products reduces exactly to an XY model of doubled angles with induced pair couplings. Complete connectivity is used below only to make those couplings uniform and obtain the Curie--Weiss solution. {
For a triangle-free graph with interactions restricted to its three-edge motifs, $\mathcal T$ is empty and this Hamiltonian is identically zero. On a graph containing triangles the induced $J^{\rm eff}_{ij}$ can be nonuniform and short ranged; the local reduction still holds, but the complete-graph transition temperature, critical exponents, and finite-size laws do not follow for such graphs. The exact mapping in $d=2$ and the Curie--Weiss solution on the complete graph are therefore separate statements.
}

Every pair belongs to exactly $N-2$ triangles. Therefore Eq. \eqref{eq:Htriad} becomes, exactly at finite $N$,
\begin{equation}
H_N=-\frac{J}{4N^2}{N\choose3}
-\frac{J(N-2)}{4N^2}\sum_{i<j}\cos\!\left[2(\theta_i-\theta_j)\right].
\label{eq:exactmapping}
\end{equation}
Introducing $\phi_i=2\theta_i$ gives a Curie--Weiss XY Hamiltonian plus an additive constant,
\begin{align}
H_N&=H_0(N)-\frac{K_N}{N}\sum_{i<j}\cos(\phi_i-\phi_j), \nonumber\\
K_N&=\frac{J(N-2)}{4N}.
\label{eq:CWXY}
\end{align}
where $K_N\to J/4$ as $N\to\infty$. Thus the apparent triadic interaction is not irreducible in the planar model: it is exactly the pairwise interaction of the nematic angles.

On the complete graph the reduction can be compressed further. From Eq. \eqref{eq:nematic},
\begin{equation}
\sum_{i<j}\cos(\phi_i-\phi_j)
=\frac12\left(N^2n^2-N\right),
\label{eq:pairsum}
\end{equation}
so the energy of every configuration is determined by $n$ alone. In the thermodynamic limit the energy per node is
\begin{equation}
    e(n)=-\frac{J}{24}-\frac{J}{8}n^2
    =-\frac{J}{24}\left(1+3n^2\right).
    \label{eq:energy_n}
\end{equation}

\subsection{Why two dimensions are special}

The reduction has a useful invariant formulation. Let the attribute vectors be unit vectors in $d$ dimensions and define the global second-moment tensor
\begin{equation}
    Q=\sum_{i=1}^N \mathbf S_i\mathbf S_i^{\mathsf T}.
\end{equation}
A direct expansion of $\Tr Q^3$ separates terms with three distinct indices from repeated-index terms and yields the exact identity
\begin{equation}
\sum_{i<j<k}
(\mathbf S_i\!\cdot\!\mathbf S_j)
(\mathbf S_j\!\cdot\!\mathbf S_k)
(\mathbf S_k\!\cdot\!\mathbf S_i)
=\frac{\Tr Q^3-3\Tr Q^2+2N}{6}.
\label{eq:Qidentity}
\end{equation}
For $d=2$, $Q$ is a $2\times2$ positive-semidefinite matrix with $\Tr Q=N$. The Cayley--Hamilton relation then gives
\begin{equation}
    \Tr Q^3=\frac{3N}{2}\Tr Q^2-\frac{N^3}{2},
    \label{eq:CH2d}
\end{equation}
so the cubic invariant in Eq. \eqref{eq:Qidentity} is not independent. This algebraic dependence is the tensor counterpart of the doubled-angle identity in Eq. \eqref{eq:triadidentity}. For $d\ge3$, $\Tr Q^3$ is generically independent of $\Tr Q^2$ and $\Tr Q$; the algebraic mechanism that enforces the planar reduction is therefore absent in general. The planar case is consequently a distinguished exactly reducible limit of the broader {continuous-attribute model}.

{
A quantitative measure of the additional invariant is obtained by writing $Q=NI_d/d+\Delta$, with $\Tr\Delta=0$. The exact Hamiltonian in dimension $d$ then decomposes as
\begin{align}
H_N^{(d)}&=H_{\rm pair}^{(d)}-\frac{J}{6N^2}\Tr\Delta^3,\label{eq:d_decomp}\\
H_{\rm pair}^{(d)}&=-\frac{J}{6N^2}\left[\frac{N^3}{d^2}-\frac{3N^2}{d}+2N
+3\left(\frac Nd-1\right)\Tr\Delta^2\right].\label{eq:d_pair}
\end{align}
Because $\Tr\Delta^2=\sum_{ij}(\mathbf S_i\!\cdot\!\mathbf S_j)^2-N^2/d$, the second expression consists only of pairwise quadrupolar interactions and a constant. For $d=2$, $\Tr\Delta^3=0$ identically. For $d=3$, the difference per node is exactly
\begin{equation}
\frac{H_N^{(3)}-H_{\rm pair}^{(3)}}{N}=-\frac J6\Tr\left(\frac QN-\frac{I_3}{3}\right)^3.
\label{eq:d_residual}
\end{equation}
For perfectly aligned three-component vectors the eigenvalues of $Q/N$ are $(1,0,0)$, so the difference is $-J/27$. Thus the independent cubic term makes a finite contribution to the ordered-state energy, even though the two-dimensional reduction is exact.
{

For a finite numerical comparison we independently sampled the three-dimensional triadic Hamiltonian and its quadratic pairwise reference at $N=128$. The same denominator $D_N=(N-1)(N-2)$ was used in both sampling Hamiltonians, as in the planar simulations described below; the formulas above refer to the $N^2$ convention. We measured the rotationally invariant three-dimensional nematic order
\begin{equation}
\mathsf Q=\frac{3Q}{2N}-\frac{I_3}{2},\qquad
q_3=\sqrt{\frac23\Tr\mathsf Q^2}.
\label{eq:q3}
\end{equation}
Each model was sampled with $28$ independent chains from each of ordered and random initial configurations, using $600$ equilibration sweeps and $260$ measurements separated by six sweeps (one sweep is $N$ single-spin proposals). The reported standard errors are determined across the $56$ chains. Table~\ref{tab:3d_check} gives the resulting order and energy per node. The differences in the low-temperature ordered state directly quantify the influence of the independent cubic interaction, while the two ensembles are much closer in the disordered regime. This limited comparison is not intended to characterize the three-dimensional phase transition.
\begin{table*}[t]
\caption{Limited comparison of independently sampled three-dimensional triadic and quadratic pair models for $N=128$, with $J=k_B=1$ and the common finite-size normalization $D_N=(N-1)(N-2)$. Statistical uncertainties (one standard error) are over independent Monte Carlo chains.}
\label{tab:3d_check}
\begin{ruledtabular}
\begin{tabular}{ccccc}
$T$ & $q_3^{\rm tri}$ & $q_3^{\rm pair}$ & $e_{\rm tri}$ & $e_{\rm pair}$\\
$0.01$ & $0.96872(11)$ & $0.95240(13)$ & $-0.156376(36)$ & $-0.118423(27)$\\
$0.12$ & $0.10693(27)$ & $0.10586(28)$ & $-0.019065(7)$ & $-0.019019(7)$\\
$0.25$ & $0.09312(26)$ & $0.09254(28)$ & $-0.018717(6)$ & $-0.018698(6)$\\
\end{tabular}
\end{ruledtabular}
\end{table*}
The configuration-level residual can also be measured without a second ensemble: in the finite-size sampling convention,
\begin{equation}
\frac{\widetilde H^{(3)}-\widetilde H_{\rm pair}^{(3)}}{N}
=-\frac{4J}{81c_N}\Tr\mathsf Q^3,\quad
c_N=\frac{(N-1)(N-2)}{N^2}.
\label{eq:d3_mc_residual}
\end{equation}
For example, at $T=0.01$ we find $\langle\Tr\mathsf Q^3\rangle=0.68179(23)$ in the triadic ensemble, corresponding to a mean residual of about $-0.0345$ per node. This within-configuration residual is distinct from the difference between independently sampled mean energies in Table~\ref{tab:3d_check}.

}
}

\subsection{Free energy and self-consistency}

Because Eq. \eqref{eq:CWXY} is the fully connected $O(2)$ model, the usual mean-field saddle point becomes exact in the thermodynamic limit \cite{stanley1968}. Choose the nematic director along the real axis so that $n=\av{\cos2\theta}$. {
In the mean-field decoupling of Eq.~\eqref{eq:CWXY}, one node interacts with the average nematic order of the others, so its effective molecular field is $h=K_Nn\to Jn/4$. This field is self-consistently generated by the interaction, not an independently applied external field. The one-node energy is $-h\cos2\theta$, and normalization over $\theta\in[0,2\pi)$ gives the angular distribution below.
} The normalized single-site distribution is
\begin{equation}
P_h(\theta)=\frac{\exp(\beta h\cos2\theta)}{2\pi I_0(\beta h)},
\qquad \beta=(k_BT)^{-1},
\end{equation}
where $I_\nu$ is a modified Bessel function. Hence
\begin{equation}
    n=\frac{I_1(\beta h)}{I_0(\beta h)}.
\end{equation}
Minimization of the variational free energy gives $h=Jn/4$ {
(equivalently, $\partial f/\partial n=(J/4)[n-I_1(\beta Jn/4)/I_0(\beta Jn/4)]=0$) 
} and therefore
\begin{equation}
    n=\frac{I_1(\beta Jn/4)}{I_0(\beta Jn/4)}.
    \label{eq:selfconsistency}
\end{equation}
Equivalently, the free-energy density is
\begin{equation}
    f(n)=-\frac{J}{24}+\frac{J}{8}n^2
    -k_BT\ln I_0\!\left(\frac{\beta Jn}{4}\right).
    \label{eq:freeenergy}
\end{equation}

\subsection{Limiting cases and critical point}

At high temperature, $I_1(x)/I_0(x)=x/2-x^3/16+O(x^5)$, so Eq. \eqref{eq:selfconsistency} gives
\begin{equation}
 n=\frac{\beta J}{8}n-\frac{(\beta J)^3}{1024}n^3+O(n^5).
 \label{eq:landauSC}
\end{equation}
The disordered solution loses stability at
\begin{equation}
    k_BT_c=\frac{J}{8}.
    \label{eq:Tc}
\end{equation}
For $T>T_c$, $n=0$ and $e=-J/24$ in the thermodynamic limit.

\begin{figure}[b]
    \centering
    \includegraphics[width=0.9\linewidth]{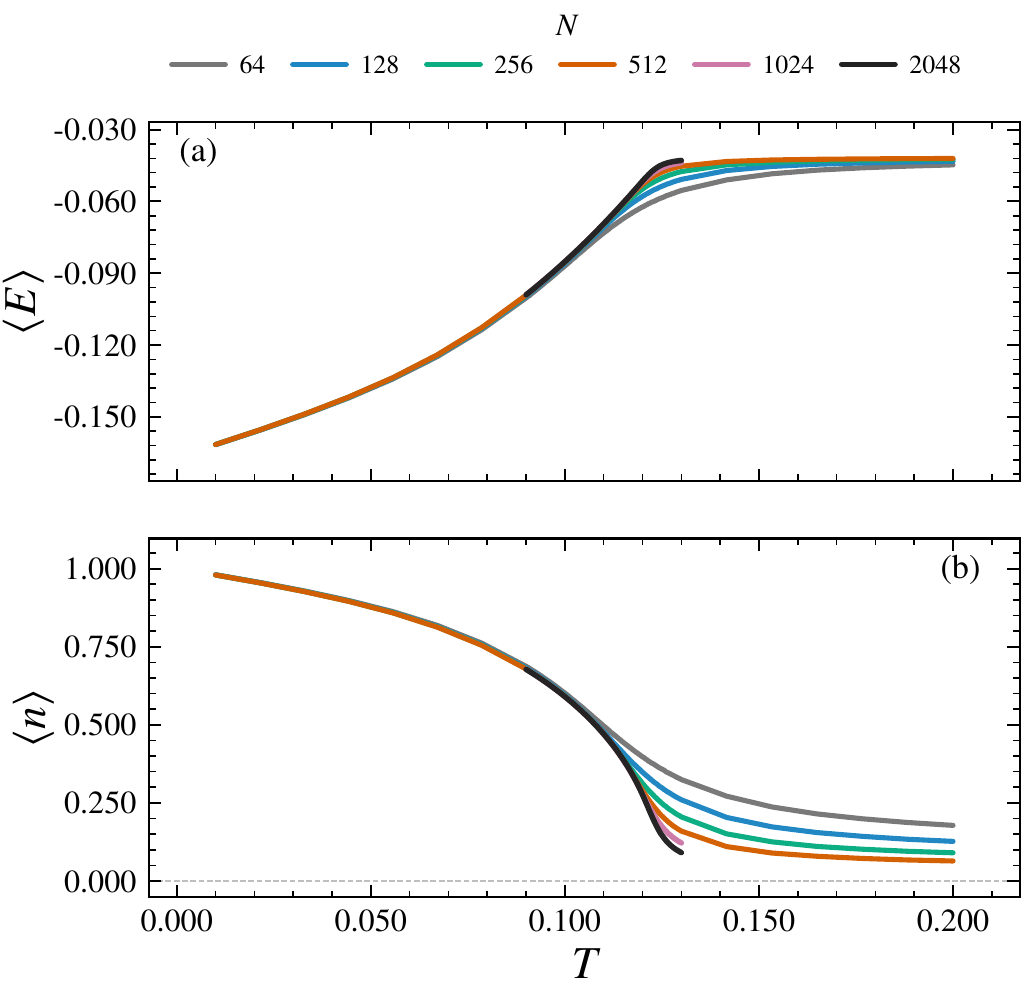}
    \caption{Temperature dependence of (a) the energy per node and (b) the nematic order parameter $n$ for $N=64$--$2048$. The curves sharpen around the exact thermodynamic critical point $T_c=1/8$ for $J=1$, while the low-temperature state approaches strong nematic order.}
    \label{fig:E_n_T}
\end{figure}

At low temperature,
\begin{equation}
    \frac{I_1(x)}{I_0(x)}
    =1-\frac{1}{2x}-\frac{1}{8x^2}+O(x^{-3}),
\end{equation}
which gives, to leading order,
\begin{equation}
    1-n\simeq\frac{2k_BT}{J},
    \qquad
    e(T\to0)=-\frac{J}{6}.
\end{equation}
{
For the Monte Carlo normalization in Eq.~\eqref{eq:MC_norm}, the fully aligned state has $e_0=-J/6\simeq-0.1667$ at every finite $N$; Fig.~\ref{fig:E_n_T}(a) approaches this value as $T\to0$.
}
Near $T_c$, with $t=(T_c-T)/T_c>0$, Eq. \eqref{eq:landauSC} yields
\begin{equation}
    n=\sqrt{2}\,t^{1/2}+O(t^{3/2}),
    \label{eq:ncritical}
\end{equation}
and therefore
\begin{equation}
    e(T)=-\frac{J}{24}-\frac{J}{4}t+O(t^2).
\end{equation}
The heat capacity per node has a finite jump,
\begin{equation}
    \Delta C=\frac{J}{4T_c}=2k_B,
\end{equation}
corresponding to $\alpha=0${
, where $\alpha$ characterizes the singular part of the heat capacity, $C_{\rm sing}\sim |T-T_c|^{-\alpha}$.
}

\subsection{Critical exponents}

{
Here $\beta$ in the following critical-exponent notation denotes the order-parameter exponent, whereas $\beta=(k_BT)^{-1}$ elsewhere denotes inverse temperature.
} Equation \eqref{eq:ncritical} gives $\beta=1/2$. To obtain the susceptibility, add a field $h_2$ conjugate to $\sum_i\cos2\theta_i$. {
This is a genuine externally imposed perturbation: the total field in the single-site equation becomes $h=Jn/4+h_2$.
} Above $T_c$,
\begin{equation}
    n\simeq\frac{\beta}{2}\left(h_2+\frac{J}{4}n\right),
\end{equation}
so
\begin{equation}
    \chi=\left.\frac{\partial n}{\partial h_2}\right|_{h_2=0}
    =\frac{\beta/2}{1-\beta J/8}
    \sim |T-T_c|^{-1},
\end{equation}
which gives $\gamma=1$. At $T=T_c$, the cubic term implies $n^3\propto h_2$, hence $\delta=3$. Together with $\alpha=0$,
\begin{equation}
    (\alpha,\beta,\gamma,\delta)=\left(0,\frac12,1,3\right),
    \label{eq:exponents}
\end{equation}
as expected for the mean-field $O(2)$ (XY) model.

\begin{figure}[b]
    \centering
    \includegraphics[width=0.9\linewidth]{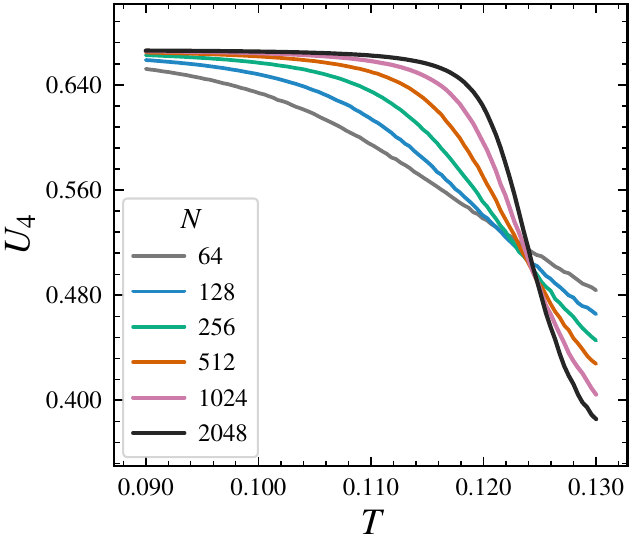}
    \caption{Binder cumulant $U_4$ as a function of temperature. The crossings drift toward the analytical critical temperature $T_c=1/8$ and occur close to the mean-field $O(2)$ critical value $U_4^*=1-\pi/6\simeq0.4764$ for the convention in Eq. \eqref{eq:binder}.}
    \label{fig:binder1}
\end{figure}

The quartic critical free energy also fixes the complete-graph finite-size powers. {
The angular measure of a two-component order parameter supplies a radial Jacobian $n$, while the critical Landau rate function begins with a positive $n^4$ term (Appendix~\ref{app:critical}).
} The typical order parameter scales as $n\sim N^{-1/4}$ and any standard nematic fluctuation susceptibility has the leading size dependence $N^{1/2}$. Thus
\begin{equation}
    n(T_c)\sim N^{-1/4},
    \qquad
    \chi(T_c)\sim N^{1/2}.
    \label{eq:FSSpred}
\end{equation}

\section{Simulation and results}

{
For all numerical results we use $k_B=1$. The Monte Carlo code samples the finite-size Hamiltonian
\begin{equation}
\begin{aligned}
\widetilde H_N&=-\frac{J}{(N-1)(N-2)}\sum_{i<j<k}\tau_{ijk}\\
&=\frac{H_N}{c_N},\qquad c_N=\frac{(N-1)(N-2)}{N^2}.
\end{aligned}
\label{eq:MC_norm}
\end{equation}
This differs from the theoretical convention in Eq.~\eqref{eq:Htriad} by an $O(N^{-1})$ normalization correction, while the thermodynamic limit is identical. Temperatures in the figures are expressed in the sampling convention. At fixed $N$ the corresponding temperature for Eq.~\eqref{eq:Htriad} is $T'=c_NT$. We simulate fully connected networks with $J=1$ and system sizes $N=64,128,256,512,1024$, and $2048$, covering both sides of $T_c=1/8$.
}

For the critical-region simulations, we use $150$ independent Monte Carlo runs at each temperature. Each run starts from the fully aligned state $\theta_i=0$ for all $i$. Equilibration consists of $9N^2$ single-agent Metropolis proposals. In each proposal an agent is chosen at random and a new angle $\theta_i'$ is drawn uniformly from $[0,2\pi)$. The move is accepted with probability
$P_{\rm acc}=\min\!\left[1,\exp(-\beta\Delta H_N)\right]$,
where {$\Delta H_N$ is the change of $\widetilde H_N$ in Eq.~\eqref{eq:MC_norm}, evaluated through its exactly reduced pairwise form.} This avoids an explicit sum over all triads for every proposed move. After equilibration, $1000$ measurement blocks are recorded, with $N^2$ attempted updates between successive measurements. The energy per node shown in Fig.~\ref{fig:E_n_T} is {$e=\widetilde H_N/N$}.

We extract the following quantities from the production phase:
\begin{itemize}
    \item Nematic Order Parameter $(n)$:
\begin{equation}
    C_2=\sum_{i=1}^N\cos2\theta_i,
    \qquad
    S_2=\sum_{i=1}^N\sin2\theta_i,
\end{equation}
through
\begin{equation}
    n=\frac{\sqrt{C_2^2+S_2^2}}{N}.
\end{equation}

    \item Specific Heat $(C_v)$:

\begin{equation}
    C_v=
    \frac{\langle \widetilde H_N^2\rangle-\langle \widetilde H_N\rangle^2}
    {NT^2}.
\end{equation}

    \item Nematic fluctuation susceptibility $(\chi)$:
\begin{equation}
    \chi=N
    \left(\langle n^2\rangle-\langle n\rangle^2\right).
    \label{eq:chiMC}
\end{equation}
Here $\chi$ denotes the dimensionless radial fluctuation observable used in the simulations. It is distinct in normalization from the field-response susceptibility used to identify $\gamma$ above, but it has the same critical size dependence $N^{1/2}$.

    \item Binder Cumulant $(U_4)$: {
We use the dimensionless ratio defined in Eq.~\eqref{eq:binder}.
}
\end{itemize}
The reported observables are averaged over $150$ independent runs, with $1000$ recorded measurements per run.

Figure \ref{fig:E_n_T} shows the continuous onset of nematic order and the corresponding change in energy. The transition region approaches $T=1/8$ as the system size increases, consistent with the analytical solution.

\subsection{Binder cumulant and critical temperature}

We use the Binder convention \cite{binder1981,fisherbarber1972}
\begin{equation}
    U_4=1-\frac{\av{n^4}}{3\av{n^2}^2}.
    \label{eq:binder}
\end{equation}
At criticality the properly scaled order-parameter distribution becomes size independent, so Binder curves for different $N$ approach a common crossing region. Figure \ref{fig:binder1} shows that the crossings move toward $T=0.125$.

For the present mean-field $O(2)$ point, the quartic radial distribution $p(n)\propto n\exp(-aNn^4)$ gives
\begin{equation}
    \frac{\av{n^4}}{\av{n^2}^2}=\frac{\pi}{2},
\end{equation}
and therefore, specifically for the convention in Eq. \eqref{eq:binder},
\begin{equation}
    U_4^*=1-\frac{\pi}{6}\simeq0.4764.
    \label{eq:BinderO2}
\end{equation}
The observed crossings near $U_4\simeq0.48$ are therefore consistent with both the exact transition temperature and the two-component nematic critical theory.

\begin{figure}[t]
    \centering
    \includegraphics[width=\linewidth]{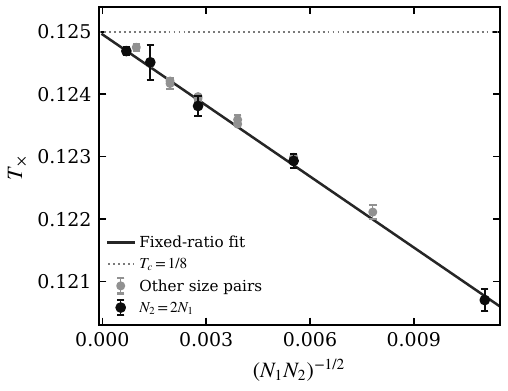}
    \caption{Binder-cumulant crossing temperatures $T_\times(N_1,N_2)$ versus $(N_1N_2)^{-1/2}$. The solid line is the weighted linear fit restricted to $N_2=2N_1$; the other size pairs provide independent visual checks. The dotted horizontal line is the exact thermodynamic value $T_c=1/8$.}
    \label{fig:fss1}
\end{figure}

{
For a pair $(N_1,N_2)$, we define $T_{\times}(N_1,N_2)$ by $U_4(T_{\times},N_1)=U_4(T_{\times},N_2)$. Close to $T_c$, the critical finite-size expansion takes the form
\begin{equation}
U_4(T,N)=U_4^*+a_1(T-T_c)N^{1/2}+a_2N^{-1/2}+\cdots.
\label{eq:binder_drift}
\end{equation}
At fixed $N_2/N_1=r>1$, cancellation of the two Binder values requires $(T_{\times}-T_c)N_1^{1/2}=O(N_1^{-1/2})$, hence $T_{\times}-T_c=O(N_1^{-1})$ (Appendix~\ref{app:critical}).
} We therefore use the fixed-ratio sequence $(64,128)$, $(128,256)$, $(256,512)$, $(512,1024)$, and $(1024,2048)$ for the quantitative extrapolation. An unweighted linear fit versus $1/N$ gives $T_c(\infty)=0.12498$, while weighting by the individual crossing uncertainties gives $T_c(\infty)=0.12496$. Both estimates converge to the exact value $T_c=1/8$. Figure~\ref{fig:fss1} also shows the remaining pairwise crossings against the symmetric inverse-size variable $(N_1N_2)^{-1/2}$ as a visual consistency check; because crossings that share a system size are statistically correlated, the all-pairs line is not used to assign the quoted critical temperature.

\begin{figure*}[t]
    \centering
    \includegraphics[width=0.8\textwidth]{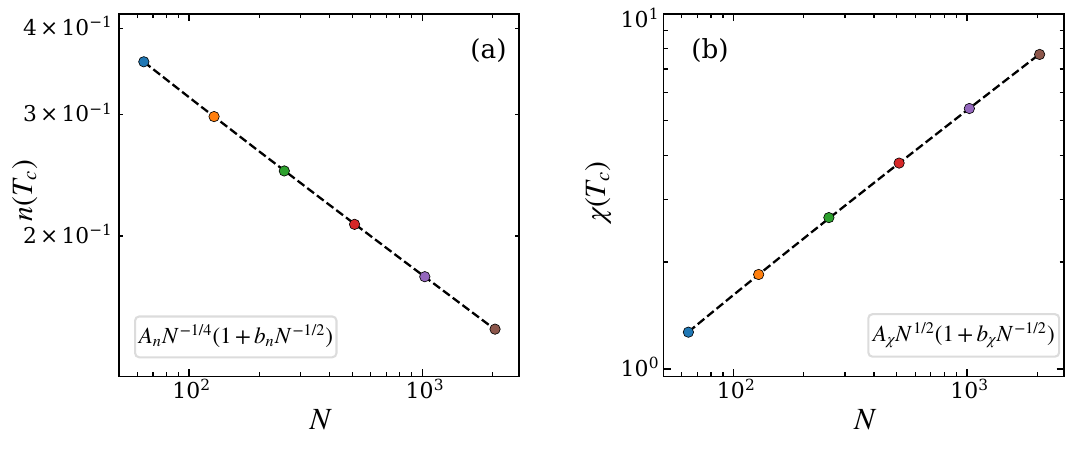}
    \caption{{Finite-size scaling at $T_c=1/8$ for $N=64$--$2048$. Points and one-standard-error uncertainties come from the tabulated Monte Carlo estimates; dashed curves are two-parameter weighted fits with fixed leading powers $-1/4$ and $1/2$ and correction power $-1/2$: (a) nematic order parameter; (b) nematic fluctuation susceptibility. Numerical coefficients are given in Table~\ref{tab:fss}.}}
    \label{fig:n_sus_T_loglog}
\end{figure*}
\subsection{Finite-size scaling}

To test the asymptotic predictions in Eq.~\eqref{eq:FSSpred}, we compare the two observables with the complete-graph scaling forms
\begin{align}
 n(T_c,N)&=A_nN^{-1/4}\left(1+b_nN^{-1/2}\right),\\
 \chi(T_c,N)&=A_\chi N^{1/2}\left(1+b_\chi N^{-1/2}\right).
 \label{eq:FSS-corrections}
\end{align}
{
The leading powers $-1/4$ and $1/2$ are exact, and the $N^{-1/2}$ correction follows from the critical finite-size expansion (Appendix~\ref{app:critical}). For each observable only $A$ and $B=Ab$ are fitted, by weighted least squares of $nN^{1/4}=A_n+B_nN^{-1/2}$ and $\chi N^{-1/2}=A_\chi+B_\chi N^{-1/2}$. Figure~\ref{fig:n_sus_T_loglog} compares the data and the resulting fits. The values in Table~\ref{tab:fss} use the six tabulated system sizes and their reported standard errors; each regression has four degrees of freedom. The fitted amplitudes approach the theoretical leading values, and the subleading coefficients quantify finite-size drift rather than independent critical exponents.
\begin{table}[b]
\caption{Two-parameter fixed-exponent fits at $T_c=1/8$, using the $N=64$--$2048$ Monte Carlo estimates. Parentheses denote one-standard-error fit uncertainties propagated from the reported error bars.}
\label{tab:fss}
\begin{ruledtabular}
\begin{tabular}{lccc}
Observable & $A$ & $b$ & $\chi^2/{\rm dof}$ \\
\hline
$n(T_c)$ & $0.97855(86)$ & $0.274(12)$ & $1.09$ \\
$\chi(T_c)$ & $0.17285(42)$ & $-0.652(31)$ & $0.99$ \\
\end{tabular}
\end{ruledtabular}
\end{table}
}

\subsection{Low-temperature bipolarity}

{
For the fully connected planar-attribute model, the exact XY reduction yields a continuous transition. Thermal binary-link balance models instead have independent link variables and can display discontinuous transitions and strong initial-condition dependence \cite{malarz2022meanfield,mohandas2025paradise}. The low-temperature equilibrium structure in the present model is determined by nematic order together with the local $\mathbb Z_2$ symmetry.
}

The local symmetry in Eq. \eqref{eq:localZ2} gives an exact interpretation of the low-temperature polar structure. In a perfectly ordered nematic state, all doubled angles satisfy $\phi_i=\phi_0$, while every original angle may independently occupy either branch,
\begin{equation}
    \theta_i=\frac{\phi_0}{2}
    \qquad\text{or}\qquad
    \theta_i=\frac{\phi_0}{2}+\pi.
\end{equation}
Hence a fixed nematic director has $2^N$ energetically equivalent polar assignments. If $N_1$ nodes occupy one branch and $N_2=N-N_1$ the opposite branch, the multiplicity is
\begin{equation}
    \Omega(N_1)={N\choose N_1}.
\end{equation}
Conditional on a given nematic configuration, the $2^N$ local $\mathbb Z_2$ sectors have equal Boltzmann weight. Hence
\begin{equation}
    P(N_1=k)=2^{-N}{N\choose k},
    \label{eq:binomial}
\end{equation}
so $\av{N_1}=N/2$ and $\mathrm{Var}(N_1)=N/4$. The all-positive link state corresponds to the two polar assignments $N_1=0$ and $N_1=N$, and therefore carries total sector weight $P_{\rm paradise}=2^{1-N}$. Paradise is thus an allowed member of the ordered manifold but is exponentially atypical in equilibrium; the overwhelmingly typical macrostate contains two nearly equal opposite groups.

{
The predominance of nearly equal-size balanced partitions is already familiar from binary structural-balance models \cite{antal1,mohandas2024multiplex}. Here its precise origin is the independent $\theta_i\mapsto\theta_i+\pi$ degeneracy of continuous-attribute configurations. The $2^N$ sectors are exactly equivalent in equilibrium; their occupation under a specific finite-time Monte Carlo trajectory is a separate dynamical question.
}

\section{Conclusion}

We studied a structural-balance Hamiltonian in which continuous pair affinities are induced by planar node attributes. The central result is an exact finite-$N$ reduction of the Heider triadic product to a pairwise Curie--Weiss XY interaction of doubled angles. The thermodynamic problem is consequently a nematic mean-field $O(2)$ model with $k_BT_c=J/8$. The same mapping fixes the complete-graph finite-size powers and the Binder-cumulant value, providing direct analytical benchmarks for the Monte Carlo data.

The ordered state retains information that is invisible in the doubled-angle description. Independent $\pi$ shifts of the original angles generate $2^N$ exactly degenerate polar assignments for every nematic configuration. Their multiplicities are binomial, so the equilibrium weight concentrates near two equally populated opposite groups, while the all-positive paradise state has exponentially small sector weight. Bipolarity therefore follows from nematic ordering together with the local $\mathbb Z_2$ degeneracy, without introducing a separate rule that enforces two camps.

Finally, the tensor identity in Eq. \eqref{eq:Qidentity} places the planar reduction in a broader setting. In two dimensions the cubic invariant $\Tr Q^3$ is constrained by Cayley--Hamilton and reduces to lower-order information; in higher dimensions it is generically independent. The present model therefore provides an exactly solvable planar benchmark for identifying when a higher-dimensional or non-mean-field extension retains genuinely independent higher-order structure.

\section{Acknowledgements}
This work was funded by the Deutsche Forschungsgemeinschaft (DFG, German Research Foundation) under Project No.~557852701 (A.A.S.).

\section{Data availability}
The numerical data and simulation code that support the findings of this article are available from the corresponding authors upon reasonable request.

\bigskip

\section*{Appendix}
\appendix
\section{Critical distribution and leading finite-size corrections}
\label{app:critical}
For the doubled-angle XY variables, the cumulant-generating function of a single uniform planar spin is $\ln I_0(z)$. Its radial large-deviation rate function is
\begin{equation}
\mathcal I(n)=nz-\ln I_0(z),\qquad n=\frac{I_1(z)}{I_0(z)},
\end{equation}
with the small-order expansion
\begin{equation}
\mathcal I(n)=n^2+\frac{n^4}{4}+\frac{5n^6}{36}+O(n^8).
\label{eq:rate}
\end{equation}
The pairwise reduction in Eq.~\eqref{eq:CWXY} contributes a Boltzmann factor $\exp[\beta K_N Nn^2/2]$, where $K_N=J(N-2)/(4N)$. At $\beta_c=8/J$, the quadratic terms cancel at leading order. The radial measure contributes a Jacobian $n$, so with $x=N^{1/4}n$ the critical density has the form
\begin{equation}
p_N(x)\propto x\,\exp(-x^4/4)\left[1+N^{-1/2}g(x)+O(N^{-1})\right],
\label{eq:scaled_radial}
\end{equation}
where $g(x)$ includes the sixth-order rate term, the finite-$N$ shift of $K_N$, and smooth prefactor corrections. Specifically, $N(5n^6/36)=N^{-1/2}(5x^6/36)$ and the finite-$N$ quadratic contribution is $2n^2=2N^{-1/2}x^2$. The leading moments consequently satisfy
\begin{equation}
\begin{aligned}
\av{n^k}&=N^{-k/4}\left[c_k+O(N^{-1/2})\right],\\
c_k&=\frac{2^{k/2}\Gamma[(k+2)/4]}{\Gamma(1/2)}.
\end{aligned}
\label{eq:critical_moments}
\end{equation}
Thus $\av n\sim N^{-1/4}$, $\chi=N(\av{n^2}-\av n^2)\sim N^{1/2}$, and both have relative corrections of order $N^{-1/2}$. In particular, $\av{n^4}/\av{n^2}^2=c_4/c_2^2=\pi/2$, yielding Eq.~\eqref{eq:BinderO2}.

For $t=(T-T_c)/T_c$ sufficiently small, the quadratic term in the exponent of the scaled order parameter is proportional to $tN^{1/2}x^2$. The cumulant therefore has an expansion of the form
\begin{equation}
\begin{aligned}
U_4(T,N)={}&U_4^*+a_1(T-T_c)N^{1/2}\\
&+a_2N^{-1/2}+O\!\left((T-T_c)^2N,N^{-1}\right).
\end{aligned}
\end{equation}
which yields a crossing shift $T_\times-T_c=O(N^{-1})$ at fixed size ratio. The same $N^{-1/2}$ corrections motivate the two-parameter forms in Eqs.~\eqref{eq:FSS-corrections}. The sampling convention in Eq.~\eqref{eq:MC_norm} changes the finite-$N$ quadratic term but not these leading powers or the correction order.

\bibliographystyle{apsrev4-2}
\bibliography{refs}

\end{document}